\documentstyle[12pt,supercite]{article}
\def\comment#1{}
\title{
\vspace{-2.5cm}
\hfill \parbox{2.3cm}{\small KOMA-96-35\\ August 1996}\\
\vspace*{1.6cm}
Convergence Behavior of Variational Perturbation Expansions}
\author{\large Wolfhard Janke$^{1}$ and Hagen Kleinert$^{2}$ \\[3mm]
\em $^1$\,Institut f\"ur Physik,  \\
\em Johannes Gutenberg-Universit\"at Mainz, \\
\em 55099 Mainz, Germany  \\
\em $^2$\,Institut f\"ur Theoretische Physik, \\
\em Freie Universit\"at Berlin, \\
\em 14195 Berlin, Germany}
\date{}
\begin{document}
\maketitle
\begin{abstract}
Variational weak-coupling perturbation theory yields
converging approximations, uniformly in the coupling strength. This allows us
to calculate directly the coefficients of {\em strong-coupling\/} 
expansions. For the anharmonic oscillator we explain the physical origin 
of the empirically observed convergence behavior which is exponentially
fast with superimposed oscillations.
\end{abstract}
%
       \section{Introduction}
%
An important problem of perturbation theory is the calculation of
physically meaningful numbers from expansions which are
usually divergent asymptotic series with coefficients growing $\propto
k!$ in high orders $k$. For small expansion parameters $g$ a direct
evaluation of the series truncated at a finite order $k \approx 1/g$ can 
yield a reasonably good approximation, but for larger couplings such 
series become completely useless and require some kind of resummation.
Well-known examples are field theoretical $\epsilon$-expansions for 
the computation of critical exponents of phase transitions, but also the 
standard Stark and Zeeman effects in atomic physics lead to divergent
perturbation expansions. 

The paradigm for studying this problem is the quantum
mechanical anharmonic oscillator with a potential
$
V(x) = \frac{1}{2} \omega^2 x^2 + \frac{1}{4} g x^4 \hspace{0.2cm}
( \omega^2,g>0).
$
The Rayleigh-Schr\"{o}dinger perturbation theory yields for the
ground-state energy a power-series expansion
\begin{equation}
   E^{(0)}(g) = \omega \sum_{k=0}^{\infty} E^{(0)}_{k} \left(
   \frac{g/4}{\omega^3} \right)^k,
\label{eq:E_exp}
\end{equation}
where the $E^{(0)}_k$ are rational numbers $1/2$, $3/4$, 
$-21/8$, $333/16$, $-30885/128$, \dots ~, which can easily be
obtained to very high orders from the recursion relations of 
Bender and Wu \cite{bewu}. Their large-order behavior is analytically
known to exhibit the typical factorial growth,
\begin{equation}
E^{(0)}_k = -(1/\pi) (6/\pi)^{1/2} (-3)^k k^{-1/2} k! (1 + {\cal O}(1/k)).
\label{eq:Ek_asy}
\end{equation}

Standard resummation methods are Pad\'e or Borel techniques whose accuracy, 
however, decreases rapidly in the strong-coupling limit. In this note
we summarize recent work on a new approach based on variational perturbation
theory \cite{syko,PI}. Our results demonstrate that by this means the 
divergent series expansion (\ref{eq:E_exp})
can be converted into a sequence of exponentially fast converging
approximations, uniformly in the coupling strength $g$
\cite{jk1,jk2,jk3,conv}. This allows us
to take all expressions directly to the strong-coupling limit, yielding
a simple scheme for calculating the coefficients $\alpha_i$ of the convergent
strong-coupling series expansion,
$E^{(0)}(g)=  (g/4)^{1/3}\left[ \alpha _0
+ \alpha _1 (4 \omega^3/g)^{2/3}
+ \alpha _2 (4 \omega^3/g)^{4/3}
+\dots\right]$.
%
       \section{Variational Perturbation Theory}
%
The origin of variational perturbation theory can be traced back to
a variational principle for the evaluation of quantum partition
functions in the path-integral formulation \cite{PI,variational}. 
While in many applications 
the accuracy was found to be excellent over a wide range of temperatures, 
slight deviations from exact or simulation results at very low temperatures
motivated a systematic study of higher-order corrections \cite{syko,PI}. 

In the zero-temperature limit the calculations simplify and lead to a 
resummation scheme for the energy eigenvalues which can be summarized as 
follows. First, the harmonic term of the potential is
split into a new harmonic term with a trial frequency $\Omega$ and a remainder,
$\omega^2 x^2 = \Omega^2 x^2 + \left(\omega^2-\Omega^2\right)x^2$, and
the potential is rewritten as
$
V(x) = \frac{1}{2} \Omega^2 x^2 + \frac{1}{4} g (-2 \sigma  x^2/ \Omega + x^4),
$
where $\sigma = \Omega ( \Omega^2 - \omega^2)/g$.
One then performs a perturbation expansion in powers of $\hat{g} \equiv
g/\Omega^3$ at a fixed $\sigma$,
\begin{equation}
\hat{E}_{N}^{(0)}(\hat g,\sigma) = \sum_{k=0}^{N} \varepsilon^{(0)}_{k}(\sigma)
\left( \hat g/4 \right)^k,
\label{eq:E_reexp}
\end{equation}
where $\hat E_N^{(0)} \equiv E_N^{(0)}/\Omega$ is the dimensionless reduced
energy. The new expansion coefficients 
$\varepsilon^{(0)}_{k}$ are easily found by inserting 
$\omega = \sqrt{\Omega^2 -g \sigma/\Omega}
= \Omega \sqrt{1 - {\hat g}\sigma}$
in (\ref{eq:E_exp}) and reexpanding in powers of $\hat g$,
\begin{equation}
   \varepsilon^{(0)}_{k}( \sigma ) = \sum_{j=0}^{k}  E^{(0)}_{j}
   \left( \begin{array}{c} (1 - 3 j)/2 \\ k-j \end{array} \right)
   (-4 \sigma )^{k-j}. 
\label{eq:eps_k}
\end{equation}
The truncated power series
$W_{N}(g,\Omega) \equiv \Omega \hat{E}^{(0)}_{N} \left(\hat{g},\sigma\right)$
is certainly independent of $\Omega$ in the limit $N \rightarrow \infty$.
At any finite order, however, it {\em does} depend on $\Omega$,
the approximation having its fastest speed of convergence where it depends 
least on $\Omega$, i.e., at points where $\partial W_N/\partial \Omega = 0$.
If we denote the order-dependent optimal value of $\Omega$ by $\Omega_{N}$,
the quantity $W_{N}(g,\Omega_{N})$ is the new approximation to $E^{(0)}(g)$.

At first sight the extremization condition $\partial W_N/\partial \Omega = 0$ 
seems to require the determination of the roots of a polynomial in
$\Omega$ of degree $3N$, separately for each value of $g$. In Ref.~\cite{jk1}
we observed, however, that this task can be greatly simplified. While
$W_N$ does depend on both $g$ and $\Omega$ separately, we could prove
that the derivative can be written as $\partial W_N/\partial \Omega=
(\hat g/4)^N P_N(\sigma)$, where
$P_N(\sigma) = -2 d \varepsilon^{(0)}_{N+1}(\sigma)/d \sigma$ is a
polynomial of degree $N$ in $\sigma$. The optimal values
of $\sigma$ were found to be well fitted by
\begin{equation}
\sigma_N = cN \left( 1 + 6.85/N^{2/3}\right),
\label{eq:sigma_opt}
\end{equation}
with $c=0.186\,047\,272\dots$ 
determined analytically (cp. Sec.~3).
This observation simplifies the calculations
considerably and shows that the optimal solutions $\Omega_N$ depend
only trivially on $g$ through $\sigma_N = \Omega_N(\Omega_N^2 -
\omega^2)/g$. Since the explicit knowledge of $\Omega_N$ is only needed 
in the final step when going back from ${\hat E}_N^{(0)}$ to $E_N^{(0)}$,
this suggests that the variational resummation scheme can be 
taken directly to the strong-coupling limit. 

To this end we introduce the reduced frequency 
$\hat{\omega} = \omega/\Omega$, write the 
approximation as
$W_N = \left( g/\hat{g} \right)^{1/3} w_N(\hat{g},\hat{\omega}^2)$,
and expand the function $w_N(\hat g,\hat \omega^2)$
in powers of $\hat{\omega}^2 = (\omega^3/g)^{2/3} \hat{g}^{2/3}$.
This gives \cite{jk2}
\begin{equation}
W_N = (g/4)^{1/3} \left[ \alpha_0
+ \alpha_1 \left(4\omega^3/g\right)^{2/3}
+ \alpha_2 \left(4\omega^3/g\right)^{4/3} +\dots \right],
\label{eq:W_N}
\end{equation}
with the coefficients,
\begin{equation}
\alpha_n = (\hat{g}/4)^{(2n-1)/3}
\sum_{k=0}^N (-1)^{k+n} \sum_{j=0}^{k-n} E_{j}^{(0)}
\left( \begin{array}{c} (1 - 3 j)/2 \\ k-j \end{array} \right)
\left( \begin{array}{c} k-j \\ n \end{array} \right)
      (-\hat{g}/4)^j.
\label{eq:alpha_n}
\end{equation}
If this is evaluated at ${\hat g} = 1/\sigma_N$ with $\sigma_N$ given in
(\ref{eq:sigma_opt}), we obtain the exponentially fast approach to the 
exact limit as shown in Fig.~\ref{fig:contour} for $\alpha_0$. The 
exponential falloff is modulated by oscillations. Our result,
$\alpha_0 = 0.667\,986\,259\,155\,777\,108\,270\,96$,
agrees to all 23 digits with the most accurate 62-digit value in the
literature. The computation of the higher-order coefficients 
$\alpha_n$ for $n>0$ proceeds similarly and the results up
to $n=22$ are given in
Table~1 of Ref.~\cite{jk2}.
%
       \section{Convergence Behavior}
%
To explain the convergence behavior \cite{jk3,conv} we recall that
the ground-state energy $E^{(0)}(g)$ satisfies a subtracted dispersion 
relation which leads to an integral representation of the original
perturbation coefficients,
\begin{equation}
E^{(0)}_k=\frac{4^k}{2{\pi}i}\int_0^{-\infty}
\frac{d g}{g^{k+1}}
\mbox{disc} E^{(0)}(g),
\label{eq:disc}
\end{equation}
where $ \mbox{disc}\,E^{(0)}(g) = 2i{\rm Im\,}E^{(0)}(g-i \eta )$
denotes the discontinuity across the left-hand cut
in the complex $g$-plane. 
For large $k$, only its $g \longrightarrow 0^-$ behavior 
is relevant and a semiclassical calculation 
yields
$\mbox{disc}\,E^{(0)}(g) \! \approx \! -2i \omega (6/\pi)^{1/2}
(-4\omega^3\!/3g)^{1/2} \exp(4\omega^3\!/3g)$, which in turn implies the
large-order behavior (\ref{eq:Ek_asy}) of $E_k^{(0)}$.

The reexpanded series (\ref{eq:E_reexp}) is obtained
from (\ref{eq:E_exp}) by the replacement of
$\omega \longrightarrow \Omega \sqrt{1- \sigma \hat g}$.
In terms of the coupling constant, the above replacement amounts to
$\bar g \equiv g/ \omega ^3 \longrightarrow \hat g/(1- \sigma \hat
g)^{3/2}$. Using this mapping it is straightforward to show~\cite{jk3}
\begin{figure}[t]
\vskip  6.0 truecm
\includegraphics{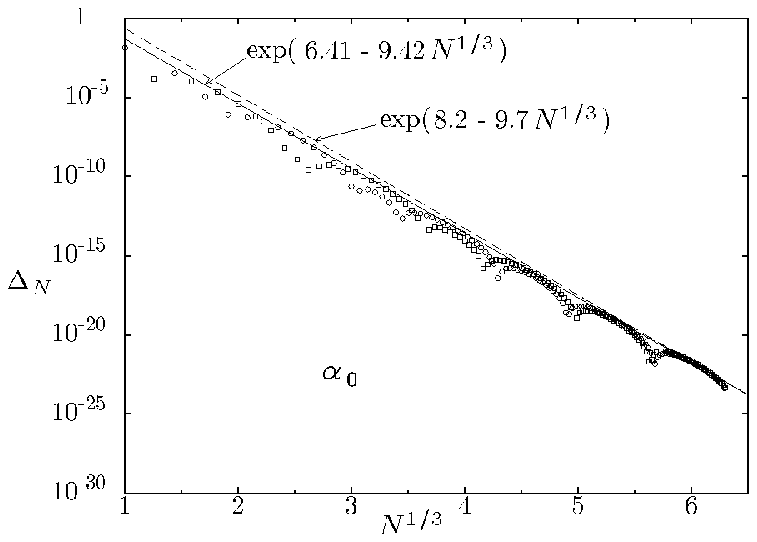}
\includegraphics{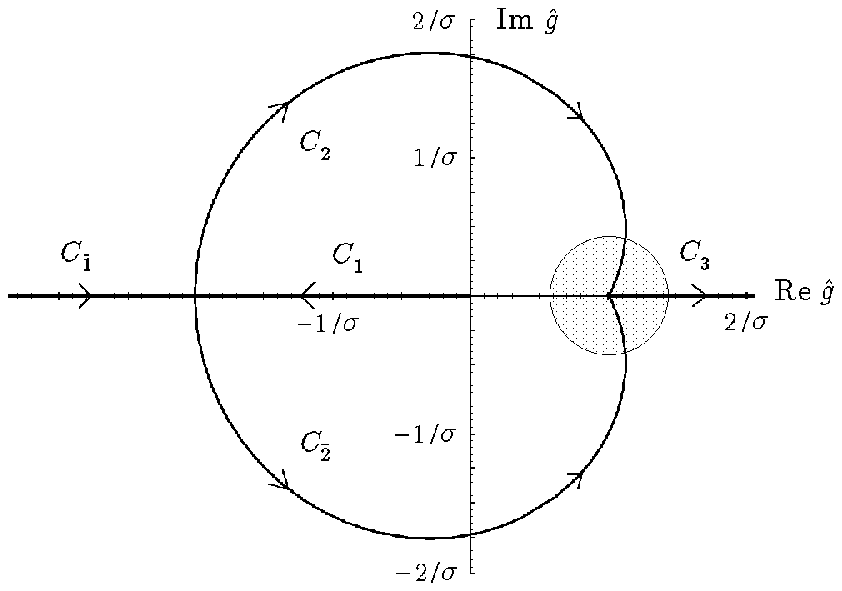}
\caption[a]{
L.h.s.: Exponentially fast convergence of the $N$th approximants for
$\alpha_0$ to the exact value. The dots show $\Delta_N =
|(\alpha_0)_N-\alpha_0|$.
R.h.s.: Cuts in the complex $\hat g$-plane. The cuts inside the shaded circle
happen to be absent due to the converge of the strong-coupling
expansion for $g > g_{\rm s}$.
}
\label{fig:contour}
\end{figure}
that ${\hat E}^{(0)} \equiv E^{(0)}/\Omega$ satisfies a dispersion 
relation in the complex $\hat g$-plane.
If $C$ denotes the cuts in this plane
and $ \mbox{disc}_C {\hat E}^{(0)}(\hat g)$
is the discontinuity across these cuts,
the dispersion integral for the expansion coefficients
$\varepsilon_k^{(0)}$ reads
\begin{equation} 
 \varepsilon^{(0)} _k=\frac{4^k}{2{\pi}i}\int_C
\frac{d\hat g}{{\hat g^{k+1}}}
\mbox{disc}_C\hat E^{(0)}(\hat g).
\label{eq:eps_disc}
\end{equation}
In the complex $\hat g$-plane, the cuts $C$ run along the contours 
$C_1,C_{\bar 1},C_2,C_{\bar 2}$, and $C_3$, as shown on the r.h.s. of 
Fig.~\ref{fig:contour}. The first four cuts are the images of the 
left-hand cut in the complex $g$-plane, and the curve $C_3$ is due to the 
square root of $1- \sigma \hat g$ in the mapping from $\bar g$ to $\hat g$.

Let us now discuss the contributions of the various cuts to the $k$th 
term $S_k$. For the cut $C_1$ and the empirically observed 
optimal solutions
$\sigma_N = cN(1+b/N^{2/3})$, a saddle-point approximation 
shows \cite{jk3} 
that this term gives a convergent contribution,
$S_N(C_1)\propto e^{ -[ -b\log(-\gamma)+(cg)^{-2/3}] N^{1/3} }$, 
only if one chooses $c=0.186\,047\,272\dots$ and
\mbox{$\gamma = -0.242\,964\,029\dots$}.
Inserting the fitted value of $b=6.85$ this yields an exponent
of
$-b\log(-\gamma) = 9.7$, 
in rough agreement with the convergence seen in Fig.~\ref{fig:contour}. 
If this was the only contribution the convergence behavior could be changed
at will by varying the parameter $b$. For $b < 6.85$, a slower convergence
was indeed observed. The convergence cannot be improved, however, by 
choosing $b > 6.85$, since the optimal convergence is limited by the 
contributions of the other cuts. 

The cut $C_{\bar 1}$ is still harmless; it contributes a last term
$S_N(C_{\bar 1})$ of the negligible order $e^{-N\log N}$.
The cuts $C_{2,\bar 2,3}$, however, deserve a careful consideration.
If they would really start
at $\hat g=1/ \sigma $, the
leading behavior would be
$\varepsilon _k^{(0)}(C_{2,\bar 2,3}) \propto \sigma ^{k}$, and therefore
$S_N(C_{2,\bar 2,3}) \propto ( \sigma \hat g)^N$,
which
would be in contradiction to the empirically observed
convergence in the strong-coupling limit. The important point is
that
the cuts in Fig.~\ref{fig:contour}
do not really reach the point
$\sigma \hat g=1$.
There exists a small circle of radius $ \Delta \hat g>0$
in which $\hat E^{(0)}(\hat g)$ has no singularities at all,
a consequence of the fact that the strong-coupling expansion
(\ref{eq:W_N}) converges for $g>g_{\rm s}$.
The complex conjugate pair of singularities
gives a contribution,
\begin{equation}
S_N(C_{2,\bar 2,3})\approx
e^{-N^{1/3}a\cos \theta}\cos(N^{1/3}a\sin \theta),
\label{eq:S_N_2}
\end{equation}
with $a=1/(|\bar{g}_{\rm s}|c)^{2/3}$. By analyzing the convergence 
behavior of the strong-coupling series we find 
$|\bar g_{\rm s}| \! \approx \! 0.160$
and $\theta \! \approx \! -0.467$, which implies for the envelope an asymptotic 
falloff of $e^{-9.23N^{1/3}}$, and furthermore also explains the
oscillations in the data~\cite{jk3}.
\section{Conclusions}
To summarize, we have shown how variational perturbation theory 
can be used to convert the divergent weak-coupling perturbation series
of the anharmonic oscillator into a sequence of converging
approximations for the strong-coupling expansion. 
By making use of dispersion relations 
and identifying the relevant singularities
we are able to explain 
the exponentially fast convergence with superimposed oscillations 
in the strong-coupling limit.\\[-0.20cm]

W.J. thanks the Deutsche Forschungsgemeinschaft for a Heisenberg
fellowship.
%
%

%
\end{document}